\documentclass{article}
\usepackage{amsmath,amssymb,exscale}
\usepackage{graphicx} 

\newcommand{\op}[1]{\mathbf #1}
\newcommand{\ket}[1]{| #1 \rangle}
\newcommand{\Ket}[1]{\pmb{\bigl|} #1 \pmb{\bigr\rangle}}
\newcommand{\bra}[1]{\langle #1 |}
\newcommand{\Eq}[1]{Eq.~(\ref{#1})}
\newcommand{\Fig}[1]{Fig.~\ref{Fig:#1}}
\newcommand{\Sec}[1]{Sec.~\ref{Sec:#1}}
\newcommand{\Id}{\mathbf 1}
\newcommand{\Mat}[2]{\left(\begin{array}{#1}#2\end{array}\right)}  
\newcommand{\mat}[1]{\left(\begin{smallmatrix}#1\end{smallmatrix}\right)}  
\newcommand{\sgm}{\boldsymbol\sigma}
\newcommand{\ro}{\boldsymbol\rho}
\newcommand{\e}{\boldsymbol e}
\newcommand{\mb}[1]{\boldsymbol #1}
\newcommand{\wlg}{without lost of generality}
\newcommand{\ie}{{\em i.e.,}}
\newcommand{\veK}{\mathbf K}
\newcommand{\opl}[1]{\mathop{#1}\limits}
\newcommand{\sep}[2]{#1 \wr #2}
\newcommand{\figframe}[1]{\centerline{\frame{~\parbox{0.95\textwidth}{
\vskip 1pt
#1
\vskip 1pt
}~}}
\vskip -4pt
}

\title{Stairway Quantum Computer}
\date{}
\author{\em Alexander Yu.\ Vlasov}
\begin{document}
\sloppy
\maketitle
\begin{abstract}
 In the paper is considered stairway-like design of quantum computer,
{\em i.e.,} array of double quantum dots or wells. The model is quite 
general to include wide variety of physical systems from coupled quantum 
dots in experiments with solid state qubits, to very complex one, like DNA 
molecule. At the same time it is concrete enough, to describe main physical 
principles for implementation of universal set of quantum gates, initialization, 
measurement, decoherence, {\em etc.} 
\end{abstract}

\section*{Introduction
\hfill\parbox[t]{4in}{
\raggedleft
\small
``Perhaps, for better understanding of this phenomenon 
[DNA replication],
we need a mathematical theory of quantum automata. Such a theory 
would provide us with mathematical models of deterministic processes 
with quite unusual properties. One reason for this is that the quantum 
state space has far greater capacity than the classical one \ldots''
\\
\medskip
Yuri I. Manin (1980)
}}

\vskip -2.8cm
\includegraphics[scale=0.33333]{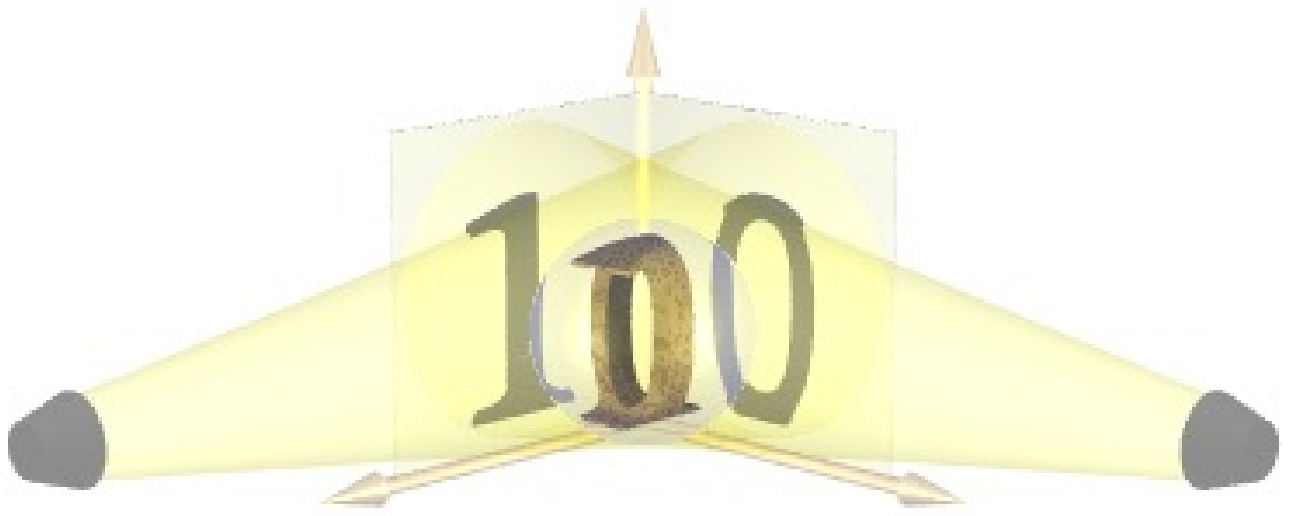} 
\vskip 1cm

\noindent
In modern theory of quantum computation is used wide range of different
models. There is class of formal mathematical models with abstract qubits.
Attempts to understand set of processes relevant to description of a real 
physical system need for other class of models with high level of complexity
and details sometime inaccessible for present level of experimental techniques.
 
The {\em stairway model} of quantum computation and control uses a compromise 
between such opposite classes.
From the one hand it is quite concrete model with array of double quantum wells, 
on the other one, the theoretical methods and mathematical equations discussed 
here are still quite general and valid for wide variety of different physical systems. 

 The model itself is introduced in \Sec{sw}. Description is general enough and 
often only basic knowledge of quantum mechanics is assumed. 
In \Sec{univ} is discussed more special theme about universal set of quantum 
gates convenient for considered model. It is set of one- and two-qubits gates 
\Eq{Enew}. Formally it uses theory of Clifford algebras and Spin groups, but 
applications may be explained with simpler methods and relevant also with some
other models in theory of quantum computing.

\section{Universality and Clifford algebras}
\label{Sec:univ}

\subsection{Universal sets of quantum gates}

A set of quantum gates is universal, if it is possible to represent 
any operation as composition of elements from the set.
Basic theory of quantum gates and networks may be found in \cite{Deu85,Deu89}. 
It is possible to use different universal sets of quantum gates.
Here it is convenient to consider a special one, based on theory of
Spin groups, Clifford and Lie algebras \cite{ClDir,Clif}. Let us 
recall it briefly.

For fixed Hamiltonian, \ie\ Hermitian matrix $\op H$, evolution of system 
is described by unitary matrix
\begin{equation}
 \op U =  \op U(\tau) = \exp(i \op H \tau),
\label{expH}
\end{equation}
where $\tau$ is time of action. 
It is possible to describe universality using the Hamiltonians \Eq{expH}
instead of quantum gates itself. It is necessary to have
some set $\op H_k$ with property: arbitrary Hamiltonian may be
expressed as linear combinations of $\op H_k$ and arbitrary number of
commutators $\op H_{k,l}=i[\op H_k,\op H_l]=i(\op H_k \op H_l - \op H_l \op H_k)$,
$\op H_{k,l,m} = i[\op H_{k,l},\op H_m]=-\bigl[[\op H_k,\op H_l],\op H_m\bigr]$, 
{\em etc.}

Usual proof of the property uses theory of Lie algebras and groups \cite{Ek95,DiVin95}
and formulae
\begin{equation}
\op U_1 \op U_2 = e^{i \op H_1 \delta}e^{i \op H_2 \delta}  
\cong e^{i (\op H_1 +\op H_2) \delta} + O(\delta^2)
\label{LieSum}  
\end{equation}
\begin{equation}
\op U_1 \op U_2 \op U_1^{-1} \op U_2^{-1} = 
e^{i \op H_1 \delta}e^{i \op H_2 \delta}
e^{-i \op H_1 \delta}e^{-i \op H_2 \delta}  
\cong e^{i (i[\op H_1,\op H_2]) \delta^2} + O(\delta^3)  
\label{LieCom}
\end{equation}
where $O(\delta^k)$ denotes order of error. 

There is some problem with practical application of expression \Eq{LieCom}. 
The term with commutator has multiplier $\delta^2$ and so it is necessary to take 
$\tau = \sqrt{\delta}$ in exponents with $\op U_1$, $\op U_2$ for construction
$\op U_{1,2}=\exp(i \op H_{1,2} \tau)$ \Eq{LieCom}. The final expression has 
error $O(\tau^{1.5})$. It was considered only first commutator, for $k$ consequent 
commutators it is necessary to take $\tau = \sqrt[k]{\delta}$, $\delta = \tau^k$.

It is not very convenient, because formulae use $\tau$ as small parameter, say
for  some expression with 10 commutators and error about $1\%$ we should use
$\delta = 10^{-10}$. It resembles ``car parking'' \cite{Clark} with almost 
mutually cancelling actions and miserable result. The construction of commutator 
\Eq{LieCom} may be more convenient for abstract proof of 
universality, than for concrete algorithms.

\subsection{Faster construction of commutators}
 
It is useful to look for some other methods of construction of ``higher order''
gates. One such method based on theory of Clifford algebras, has 
first order term $\delta$ instead of $\delta^2$ and {\em zero error} \cite{Clif}.
Let us consider special choise of Hamiltonians, when for any two of them
is true formulae
\begin{equation}
 \op H_k \op H_j = \pm \op H_j \op H_k,
\quad \op H_k^2 = \Id.
\label{pmCom}
\end{equation}
where $\Id $ is unit matrix.
The case $\op H_k \op H_j = \op H_j \op H_k$ is trivial (commutator simply is zero)
and for $\op H_k \op H_j = -\op H_j \op H_k$
instead of \Eq{LieCom} it is possible to use
\begin{equation}
e^{i\tau\,i[\op H_k,\op H_j]/2} = e^{-\tau\op H_k\op H_j} =
e^{i\frac{\pi}{4}{\op H_k}} e^{i\tau \op H_j} e^{-i\frac{\pi}{4}{\op H_k}}.
\label{ClifCom} 
\end{equation}
Really, 
$$
e^{i\frac{\pi}{4}{\op H_k}} e^{i\tau \op H_j} e^{-i\frac{\pi}{4}{\op H_k}} =
\exp\bigl(e^{i\frac{\pi}{4}{\op H_k}} i\tau \op H_j e^{-i\frac{\pi}{4}{\op H_k}}\bigr),
$$ 
but for any operator with property $\op H^2 = \Id$ it can be written
$$
e^{i \phi \op H} = \cos(\phi)\Id + i \sin(\phi) \op H,
\quad \mbox{and so,} \quad
e^{\pm i\frac{\pi}{4}{\op H_k}} = \tfrac{\sqrt{2}}{2}(\Id \pm i \op H_k),
$$
finally,
$$
\tfrac{\sqrt{2}}{2}(\Id + i \op H_k) \op H_j \tfrac{\sqrt{2}}{2}(\Id - i \op H_k) =
\tfrac{1}{2}(\Id + i \op H_k) (\Id + i \op H_k) \op H_j =
i\op H_k \op H_j.
$$

\subsection{Universality, Pauli matrices, and Spin groups}

For quantum computation with $n$ two-dimensional systems (qubits) a simple set 
of Hamiltonians satisfying \Eq{pmCom} may be expressed as tensor products of 
Pauli matrices
\begin{equation}
 \sgm_{i_1} \otimes \sgm_{i_2} \otimes \cdots \otimes \sgm_{i_n},
\label{psigm}
\end{equation}
there $\sgm_{i_k}$ are either one $\sgm_0=\Id=\mat{1&0\\0&1}$ or Pauli matrices
\begin{equation}
 \sgm_1=\sgm_x = \mat{0&1\\1&0}, \quad
 \sgm_2=\sgm_y = \mat{0&-i\\i&0}, \quad
 \sgm_3=\sgm_z = \mat{1&0\\0&-1}
\end{equation}
 
More abstract example is {\em Clifford algebra} with generators $\e_k$ satisfying
relations
\begin{equation}
 \{\e_k,\e_j\} = \e_k\e_j + \e_j\e_k = 2\delta_{kj}.
\end{equation}
Both cases are close related and for Clifford algebra with $2n$ generators
there is well known representation \cite{ClDir} using products like \Eq{psigm}
\begin{eqnarray}
 \e_{2k} & = &
  {\underbrace{\sgm_z\otimes\cdots\otimes \sgm_z}_k\,}\otimes
 \sgm_x\otimes\underbrace{\Id\otimes\cdots\otimes\Id}_{n-k-1} \, ,
 \nonumber\\
 \e_{2k+1} & = &
 {\underbrace{\sgm_z\otimes\cdots\otimes \sgm_z}_k\,}\otimes
 \sgm_y\otimes\underbrace{\Id\otimes\cdots\otimes\Id}_{n-k-1} \, .
 \label{defE}
\end{eqnarray}

System with $n$ qubits is described by $2^n$-dimensional Hilbert space,
group of unitary operators acting on such space is U$(2^n)$. Dimension
of the group is $4^n$. The same dimension has linear space of all
possible Hamiltonians, \ie\ Hermitian $2^n \times 2^n$ matrices
and as basis of the space may be used $4^n$ different {\em tensor products}
of four matrices $\sgm_k$ \Eq{psigm}. 

Any such Hamiltonian may be also expressed as {\em product} of $2n$ elements 
\Eq{defE}, but in condition of universality are used {\em commutators}.
Let us recall some facts \cite{Clif} about the set \Eq{defE}. The set 
is not universal, such Hamiltonians generate only some 
subgroup of SU$(2^n)$. The subgroup is isomorphic with Spin$(2n+1)$
and so only $(2n+1)n$-dimensional\footnote{Spin$(k)$ group is 2-1 cover
of group SO$(k)$ of rotations in $k$-dimensional space and has the same 
dimension $k(k-1)/2$.} \cite{Clif}. 

On the other hand, it is enough to annex only one Hamiltonian, say 
\begin{equation}
 \mb{g}_2 = \e_0\e_1\e_2 = \sgm_x^{(2)} = 
 \Id\otimes\sgm_x\otimes\underbrace{\Id\otimes\cdots\otimes\Id}_{n-2}\, , 
 \label{E3}
\end{equation}
to produce universal set.

It is also convenient instead of $\e_k$ \Eq{defE} to use products
$\mb{d}_k \equiv i\e_k\e_{k+1}$ together with $\e_0$, 
because they correspond to one- and two-qubits gates
\begin{eqnarray}
 &\e_0  = \sgm_x^{(1)} = 
 \sgm_x\otimes\underbrace{\Id\otimes\cdots\otimes\Id}_{n-1}\, , 
 \label{E0} \\
 & \mb{d}_{2k} = \sgm_z^{(k+1)} =
  {\underbrace{\Id\otimes\cdots\otimes \Id}_k\,}\otimes
 \sgm_z\otimes\underbrace{\Id\otimes\cdots\otimes \Id}_{n-k-1}\, , 
 \label{E2}\\
 & \mb{d}_{2k+1} = \sgm_x^{(k+1)}\otimes\sgm_x^{(k+2)} =
  {\underbrace{\Id\otimes\cdots\otimes \Id}_k\,}\otimes
 \sgm_x\otimes\sgm_x\otimes\underbrace{\Id\otimes\cdots\otimes \Id}_{n-k-2}\, . 
 \label{E22}
\end{eqnarray}
So we have $2n+1$ Hamiltonians described by equations 
Eqs.~(\ref{E3}--\ref{E22})
and they are universal, \ie\ may generate whole $4^n$-dimensional
space of all possible operations for system with $n$ qubits.
 
The set of gates may be divided on two parts: first one is $n$ one-qubit gates 
\Eq{E2} together with $n-1$ two-qubit gates \Eq{E22}, they
generate subgroup isomorphic to Spin$(2n)$. 
Second part, the two extra gates \Eq{E0} and \Eq{E3}
makes the set universal.  

\subsection{Model with spin-1/2 systems}

\begin{figure}[htb]
\figframe{
\begin{center}
\includegraphics{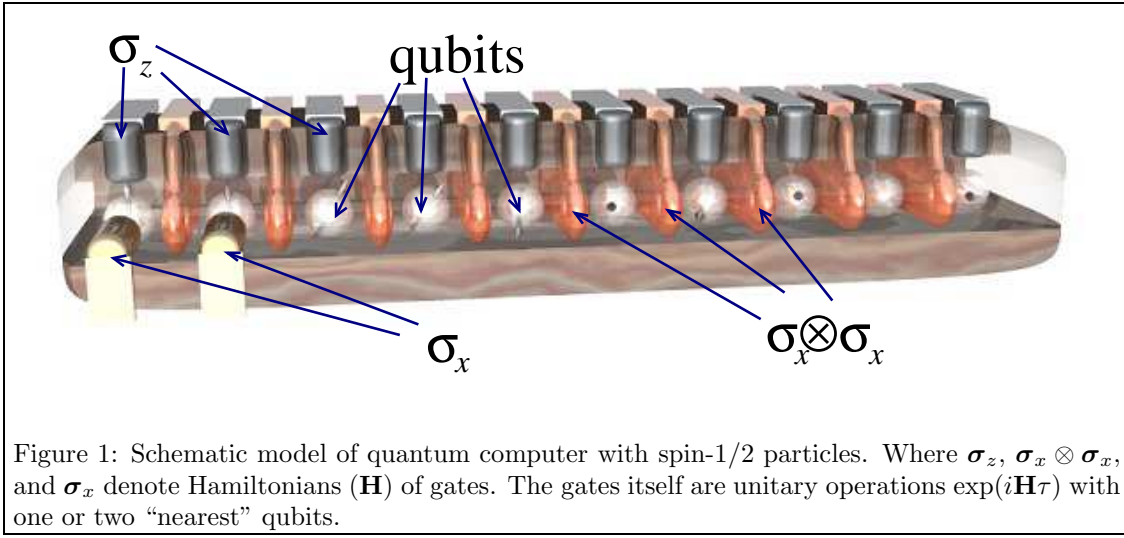} 
\end{center}
\caption{Schematic model of quantum computer with spin-1/2 particles.
Where $\sgm_z$, $\sgm_x \otimes \sgm_x$, and $\sgm_x$ denote Hamiltonians 
($\op H$) of gates. The gates itself are unitary operations $\exp(i \op H \tau)$
with one or two ``nearest'' qubits.}
\label{Fig:spins}
}
\end{figure}

\begin{figure}[htb]
\figframe{
\begin{center}
a)\includegraphics[scale=0.5]{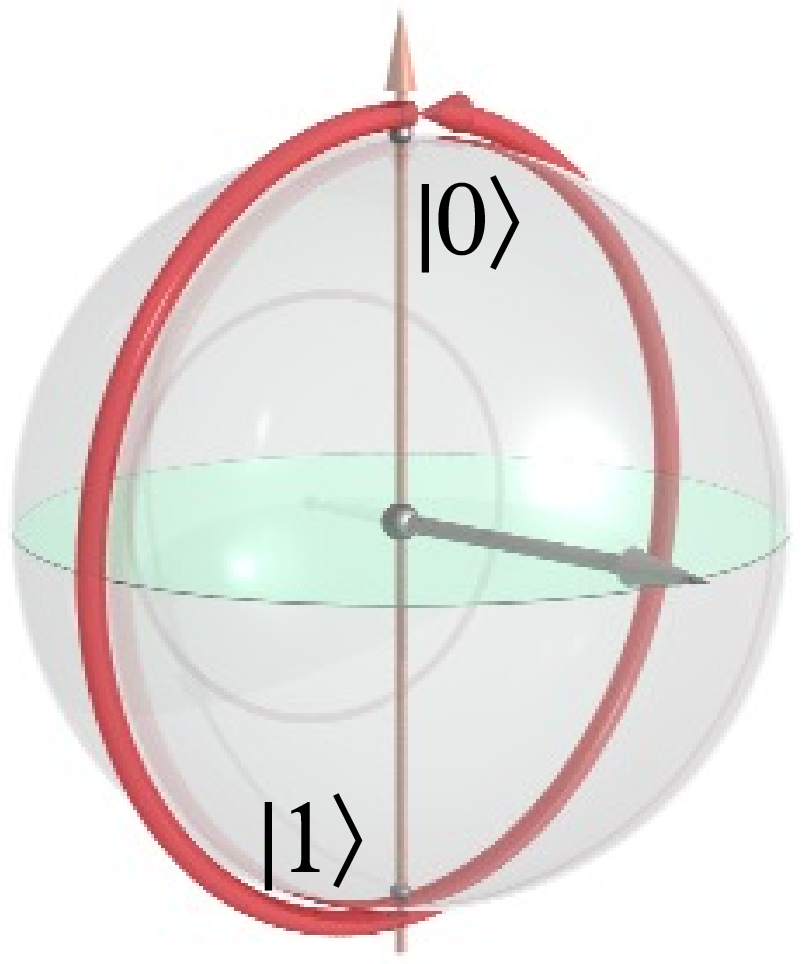} 
~
b)\includegraphics[scale=0.5]{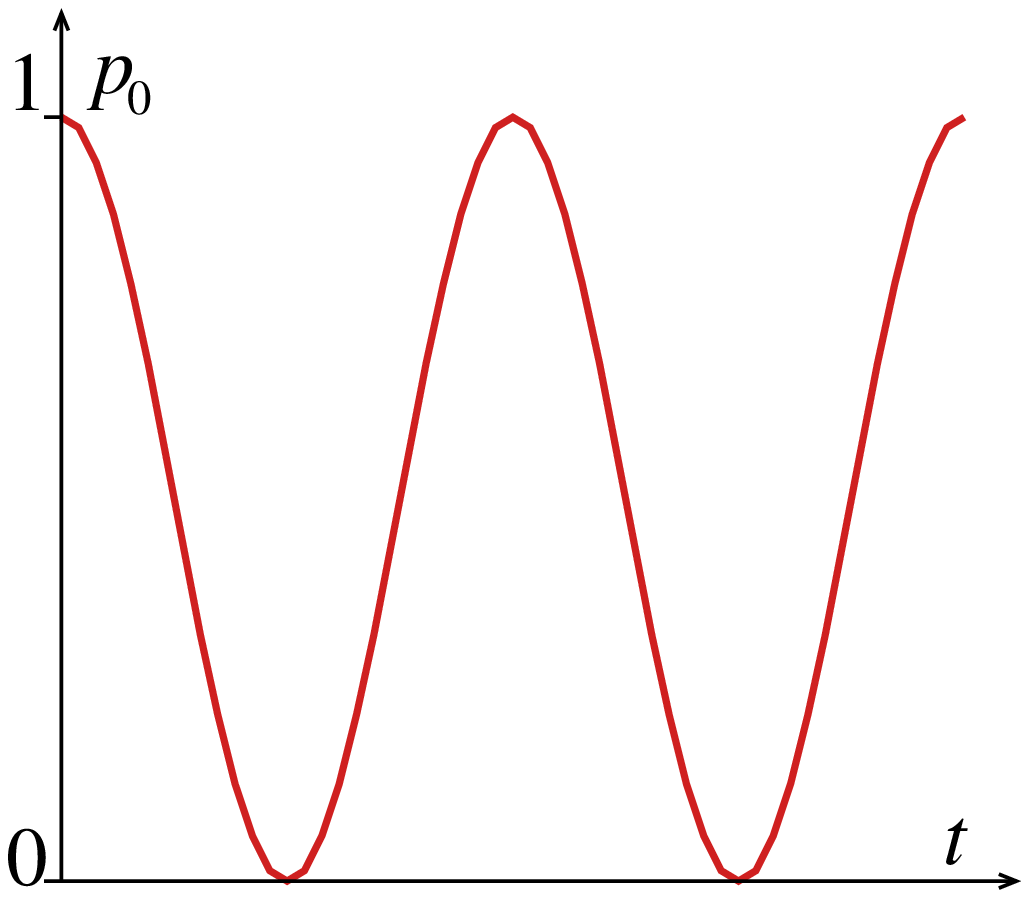}
\end{center}
\caption{Rotation around $x$ axis used for representation
of Hamiltonian $\sgm_x$.}\label{Fig:rotx}
}
\end{figure}

An universal quantum computer based on such set of gates is depicted on \Fig{spins}. 
This model with spin-1/2 particles is rather abstract, but convenient
as an understanding scheme. For example $\sgm_z$ and $\sgm_x$ may be represented
as rotation of ``spin direction'' (Bloch vector) around $z$ and $x$ axis respectively 
\Fig{rotx}.

The visual models like \Fig{rotx} are widely used for spin-1/2 system in NMR 
quantum information processing \cite{nmr} and really have universal application 
for any two-states quantum system (qubit).

Any Hamiltonian for such system is $2 \times 2$ Hemitian matrix
and may be represented in form
\begin{equation}
 \op H = h_0 \Id + h_1 \sgm_x + h_2 \sgm_y + h_3 \sgm_z
\label{QtH}
\end{equation}
where $h_k$ are real numbers. The density matrix of such system 
also may be presented in similar form
\begin{equation}
 \ro = (\Id + r_x \sgm_x + r_y \sgm_y + r_z \sgm_z)/2,
\label{Qtro}
\end{equation}
where $r_\nu$ again are real numbers. 
For pure state the vector $r$ has unit length
\begin{equation}
\ro = \ket{\psi}\bra{\psi},
\quad \ro^2 = \ro,
\quad
 r_x^2 + r_y^2 + r_z^2 = 1.
\end{equation}
Action of Hamiltonian is
\begin{equation}
 \ket{\psi(t)} = e^{-i t \op H} \ket{\psi},
 \quad
 \ro(t) = e^{-i \op H t} \ro e^{i \op H t},
\label{Hro}
\end{equation}
but \Eq{Hro} rewritten with $h_k$ \Eq{QtH} and $r_\nu$ \Eq{Qtro} corresponds 
to rotation of vector $(r_x,r_y,r_z)$ around axis $(h_1,h_2,h_3)$ in some
abstract 3D space due to usual properties of Pauli matrices%
\footnote{Three matrices $i\sigma_\nu$ coincide
with definition of {\em quaternions} and so it is representation
of 3D rotations by quaternions well known in mathematics since Hamilton.}.

For given density matrix $\ro$ usual formula for probabilities associated with
states $\ket{0}$ and $\ket{1}$ may be also expressed using \Eq{Qtro}
\begin{equation}
 p_0 = \bra{0}\ro\ket{0} = \frac{1+r_z}{2}, \quad
 p_1 = \bra{1}\ro\ket{1} = \frac{1-r_z}{2},
\end{equation}
where $r_z$ is coordinate on Bloch sphere. Well known oscillatory
behavior of such probability for Hamiltonian $\sgm_x$ is
shown on \Fig{rotx}b.

Implementation of two-qubit gates is less obvious, say Hamiltonian 
$\sgm_x\otimes\sgm_x$ corresponds to Heisenberg-like interaction
(anisotropic along $x$ axis).
 
Rest part of the work uses model with quantum wells instead
of spin-half systems.
It is more convenient also to exchange $\sgm_x$ and $\sgm_z$ in all definitions
above and use set of gates
\begin{equation}
\e'_0  = \sgm_z^{(1)},\quad \mb{g}'_2  = \sgm_z^{(2)}, \quad
\mb{d}'_{2k} = \sgm_x^{(k+1)}, \quad
\mb{d}'_{2k+1} = \sgm_z^{(k+1)}\otimes\sgm_z^{(k+2)}
\label{Enew}
\end{equation}

\section{Stairway-like model of quantum gates array}
\label{Sec:sw}

\subsection{Double quantum wells (dots) model}

\begin{figure}[htb]
\figframe{
\begin{center}
\includegraphics{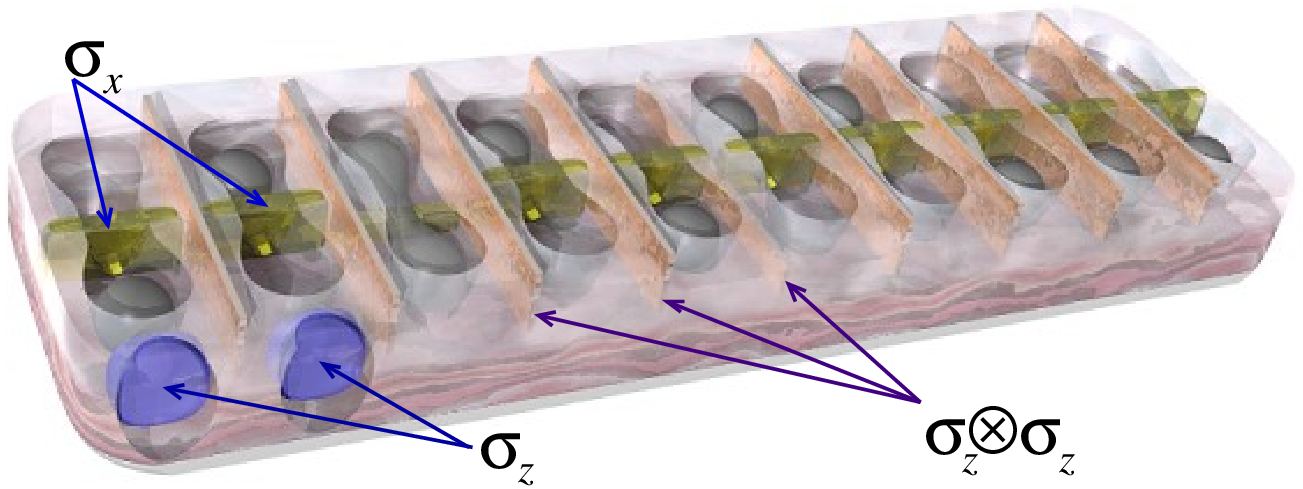} 
\end{center}
\caption{Array of double quantum wells.}\label{Fig:qudots}
}
\end{figure}

Despite of using theory of Spin groups, set of gates described above has
wider area of applications, than spin-1/2 quantum systems. Here is 
discussed design with quantum wells, where all gates may be described using 
elementary methods. It is used modified set of gates \Eq{Enew}.  
The model with array of quantum wells (sometime also called
double quantum dots) is depicted on \Fig{qudots}.

\begin{figure}[htb]
\figframe{
\begin{center}
a)\includegraphics[scale=0.5]{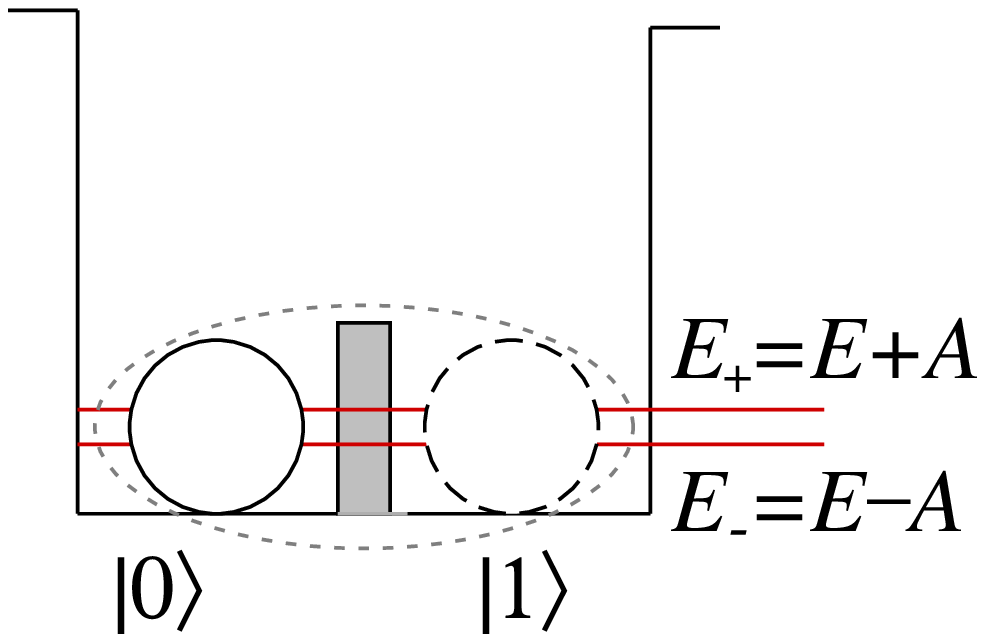}~
b)\includegraphics[scale=0.5]{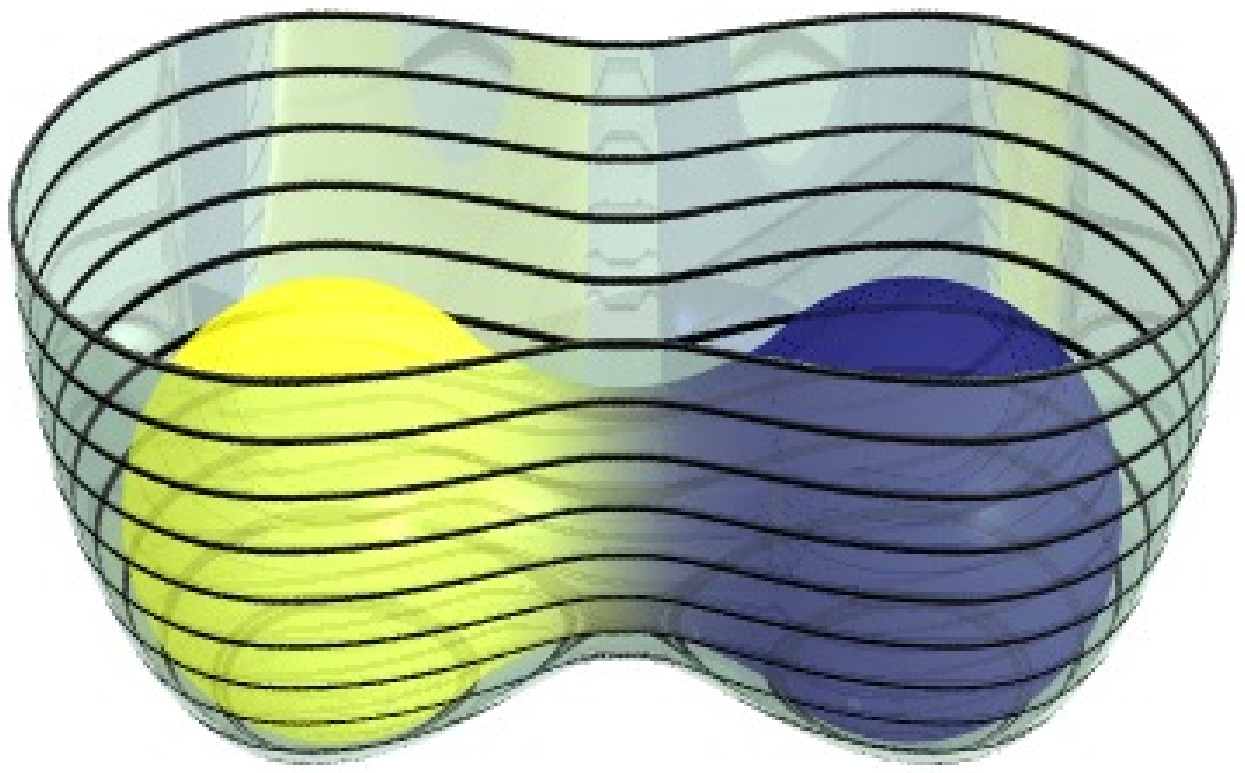} 
\end{center}
\caption{Double quantum well. As ``computational'' basis $\ket{0}$, $\ket{1}$ 
are used two states localized in first and second part of well respectively 
(formally such description more instructive for case depicted on \Fig{welld}
with bigger potential barrier and small amplitude of transition).
Eigenstates may be expressed as $\ket{\pm}=\ket{1}\pm\ket{0})/\sqrt{2}$.
a) Scheme with rectangular partitions. 
b) 3D diagram of potential and state $\ket{-}=(\ket{1}-\ket{0})/\sqrt{2}$.}
\label{Fig:wellb}
}
\end{figure}

\begin{figure}[htb]
\figframe{
\begin{center}
a)\includegraphics[scale=0.5]{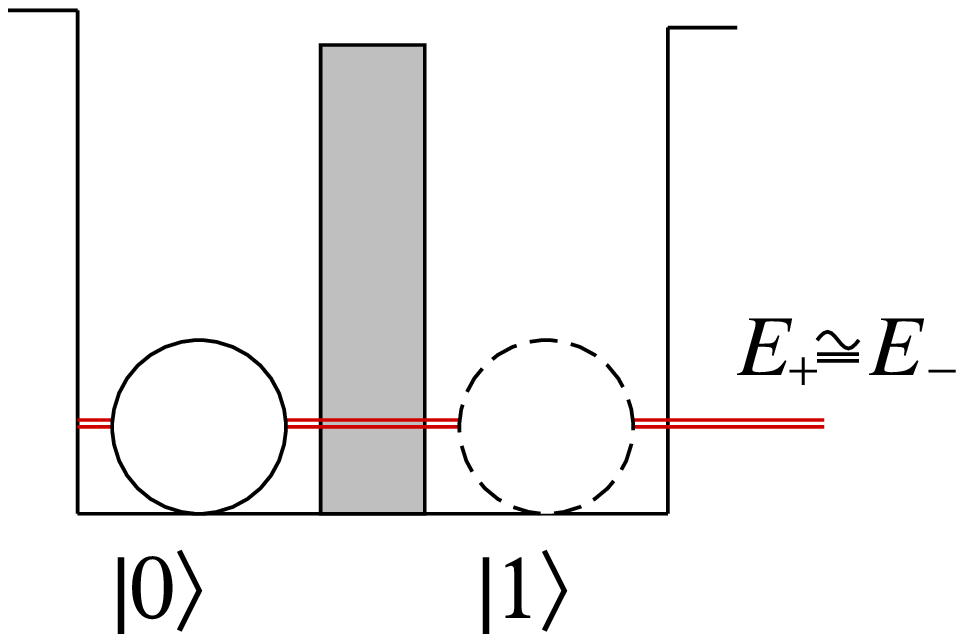}~
b)\includegraphics[scale=0.5]{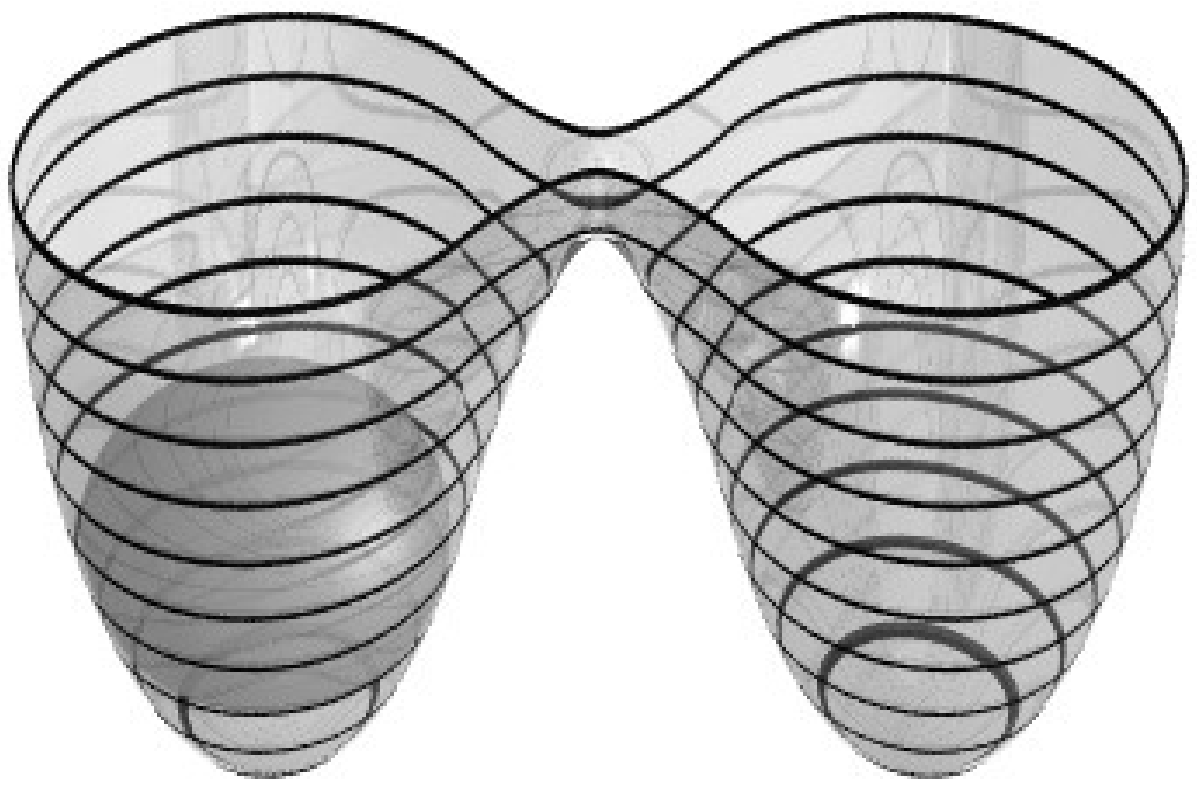} 
\end{center}
\caption{Potential with negligible tunneling amplitude.}\label{Fig:welld}
}
\end{figure}

Diagrams of such quantum wells for different potentials are shown on 
\Fig{wellb} and \Fig{welld}.
 
For second case, \Fig{welld}, it is possible to talk practically about 
two different wells with exponentially small amplitude of tunelling 
and denote state of system in first or second one as $\ket{0}$ and $\ket{1}$ 
respectively.

Let us wrote a Hamiltonian in the computational basis as
\begin{equation}
 \op H = \Mat{ll}{H_{00} &  H_{01} \\ H_{10} &  H_{11}},
\quad H_{01} = \bar{H}_{10}.
\end{equation}
Two eigenvalues, \ie\ energies of stationary states may be 
expressed as \cite{FLPQ}
\begin{equation}
 E_{\pm} = \frac{H_{00}+H_{11}}{2}\pm 
 \sqrt{\Bigl(\frac{H_{00}-H_{11}}{2}\Bigr)^2 + H_{01}H_{10}}.
\label{Epm}
\end{equation}
In considered case $H_{00}=H_{11}=E$ due to symmetry and 
using notation $A_T= |H_{01}|$ for transition (tunneling) term, it is
possible to rewrite \Eq{Epm} as \cite{FLPQ}
\begin{equation}
E_{\pm} = E \pm A_T.
\label{Epms}
\end{equation}
In such a case corresponding eigenvectors are
\begin{equation}
 \ket{+} = \frac{\sqrt{2}}{2}(\ket{1}+\ket{0}),
\quad
 \ket{-} = \frac{\sqrt{2}}{2}(\ket{1}-\ket{0}).
\label{ketpm}
\end{equation}

The value $A_T$ may be arbitrary small if two positions of the system
on \Fig{welld} are separated enough and so we may have almost degenerate
system $E_+ \cong E_- \cong E$ convenient for safe keeping state of such qubit.

If $H_{00}\neq H_{11}$  eigenvectors may be described as \cite{FLPQ}
\begin{equation}
\ket{\pm} = c_0^\pm\ket{0}+c_1^\pm\ket{1},
\quad c_0^\pm = \frac{H_{01}}{D_\pm},
\quad c_1^\pm = \frac{E_\pm - H_{00}}{D_\pm},
\quad D_\pm = e^{i\phi}\sqrt{(E_\pm-H_{00})^2+ H_{01}H_{10}}.  
\end{equation}

Finally, wave function of arbitrary state (nonstationary solution) is 
\begin{equation}
 \ket{\psi} = \ket{+}e^{-i t E_+/\hbar}+\ket{-}e^{-i t E_-/\hbar} 
 = \ket{+}e^{-i \omega_+ t}+\ket{-}e^{-i \omega_- t },
\label{nonst}
\end{equation}
so if we initially have well-separated parts \Fig{welld} and system in
state $\ket{0}$, then the change of coupling parameter $A_T$ to some nonzero 
value \Fig{wellb} is producing oscillations between $\ket{0}$ and $\ket{1}$. 

Really, $\ket{0} = (\ket{+}-\ket{-})/\sqrt{2}$, but
terms of the expression \Eq{nonst} have different frequencies $\omega_\pm$ 
and so after time $T_{1/2}$, where $(\omega_+ - \omega_-)T_{1/2} = \pi/2$, 
the system has state $\ket{1} = (\ket{+}+\ket{-})/\sqrt{2}$.
Such operation with different durations $\tau$ generates quantum gates 
corresponding to Hamiltonian $\sgm_x$. 

\begin{figure}[ht]
\figframe{
\begin{center}
a)\includegraphics[scale=0.5]{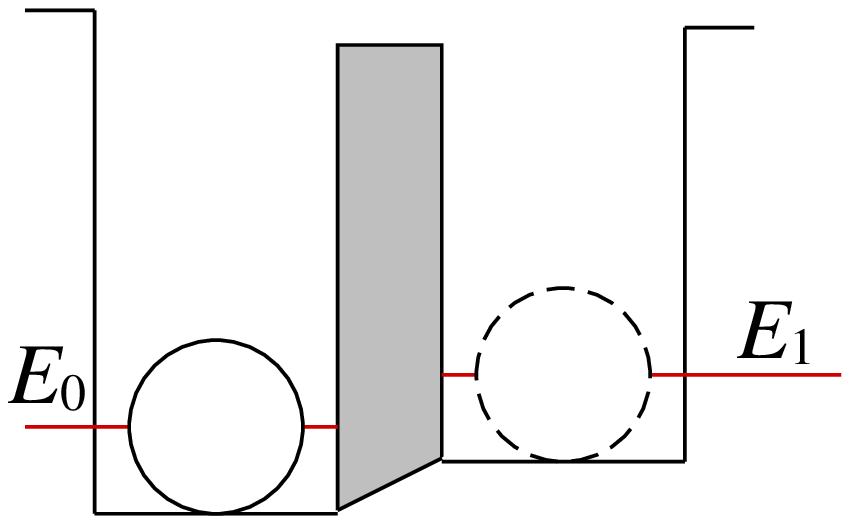}~
b)\includegraphics[scale=0.5]{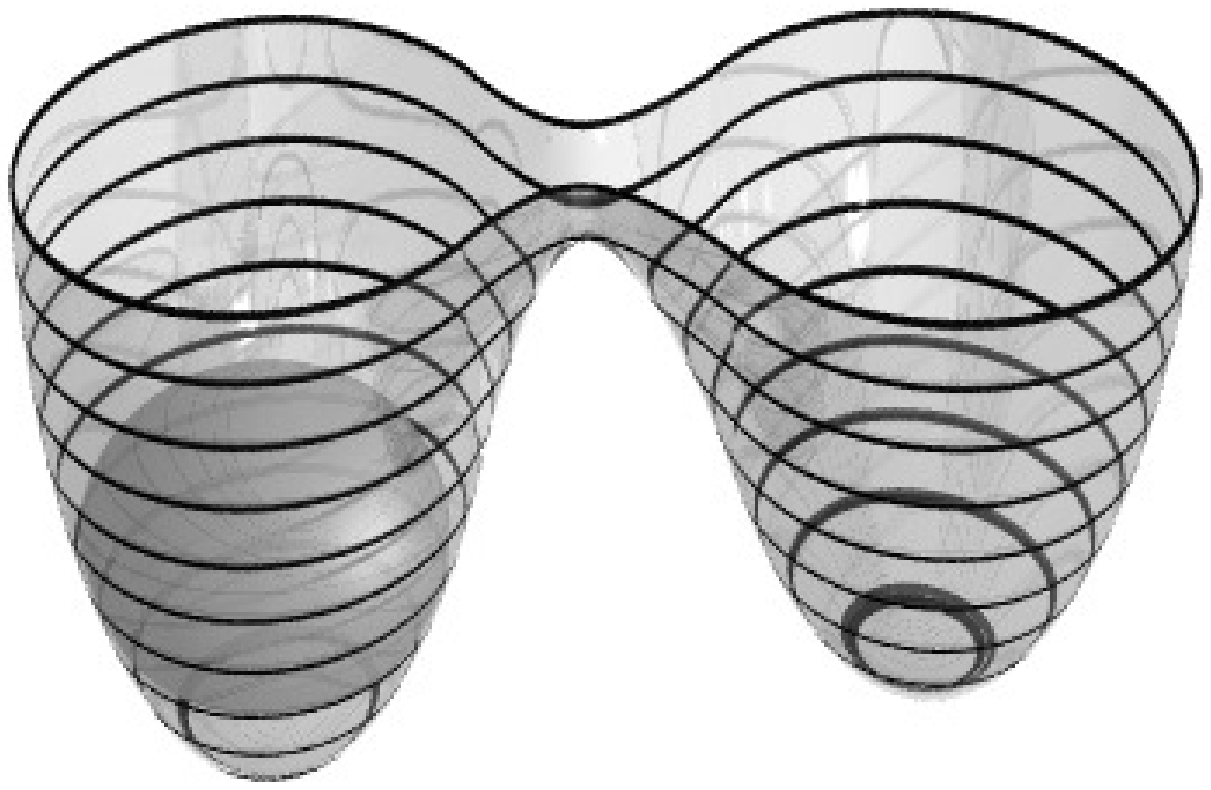} 
\end{center}
\caption{Asymmetric potential.}\label{Fig:wella}
}
\end{figure}

Distortion of potential \Fig{wella}
$H_{00} \ne H_{11}$ let us implement $\sgm_z$.

It is possible directly describe all the one-qubit gates using decomposition
of Hamiltonian on Pauli matrices $\op H = \sum_k h_k \sgm_k$
\Eq{QtH} with coefficients
\begin{equation}
 h_0 = \frac{H_{00}+H_{11}}{2}, \quad
 h_1 = \frac{H_{01}+H_{10}}{2}, \quad
 h_2 = \frac{H_{01}-H_{10}}{2i}, \quad
 h_3 = \frac{H_{00}-H_{11}}{2},
\end{equation}
where $h_1$ and $h_2$ are real and complex part of transition amplitude.
It is also convenient to write \Eq{QtH} in normalized form
\begin{equation}
 \op H = h_0 \Id + h_r (h_x \sgm_x + h_y \sgm_y + h_z \sgm_z) 
\quad h_r = \sqrt{h_1^2+h_2^2+h_3^2},
~h_x=\frac{h_1}{h_r},~h_y=\frac{h_2}{h_r},~h_z=\frac{h_3}{h_r},
\end{equation}
and in such a case 
\begin{equation}
e^{i\op H t} = e^{i h_0 t}\bigl(\cos(h_r t)\Id 
 + i \sin(h_r t)(h_x \sgm_x + h_y \sgm_y + h_z \sgm_z)\bigr),
\quad
(h_x^2+h_y^2+h_z^2)=1.
\end{equation}
 
Formally for complex $H_{01}$ instead of $\sgm_x$ we are applying
some combination of $\sgm_x$ and $\sgm_y$, but it does not
matter, because all expressions used above in discussion about
universality may be rewritten for substitutions like 
$\sgm_x \to h_x \sgm_x + h_y \sgm_y$, $\sgm_y \to h_x \sgm_y - h_y \sgm_x$
and similar property was already used ($\sgm_x\leftrightarrow\sgm_z$) 
in \Eq{Enew}. So it is possible to consider Hamiltonians with 
real $H_{01}$ \wlg. 

It was already discussed that understanding model with Bloch sphere
may be used for any two-states quantum system due to representation
of density matrix in form \Eq{Qtro}. Application of $\sgm_x$ Hamiltonian 
was depicted on \Fig{rotx} above (page \pageref{Fig:rotx}).

\begin{figure}[htb]
\figframe{
\begin{center}
a)\includegraphics[scale=0.5]{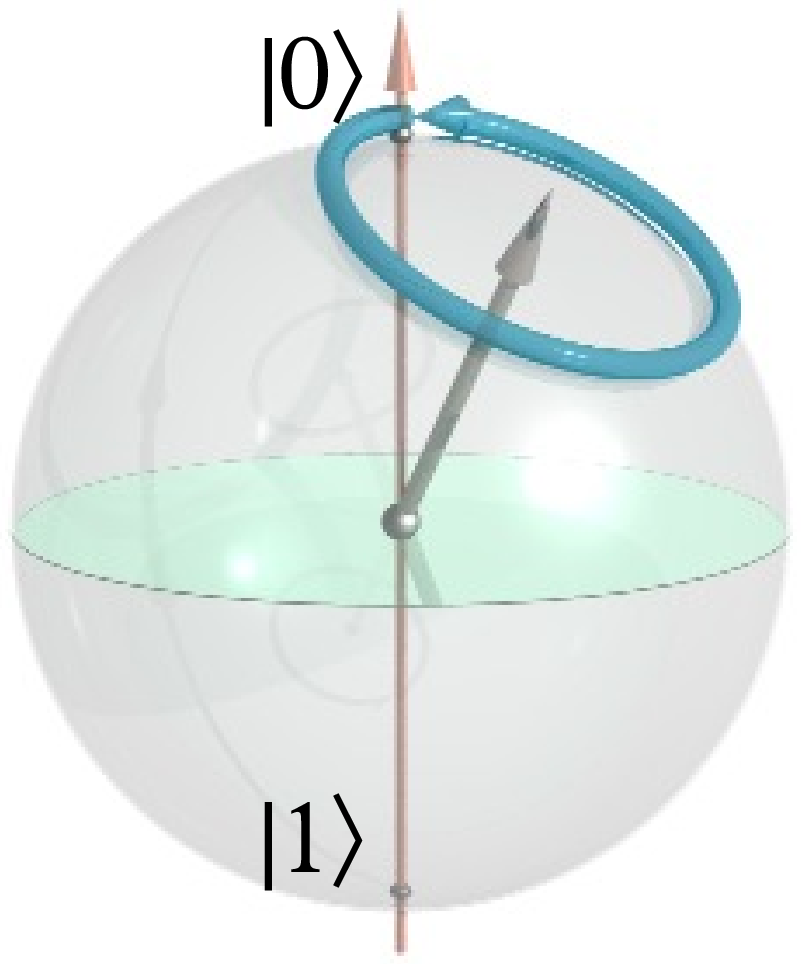} 
~
b)\includegraphics[scale=0.5]{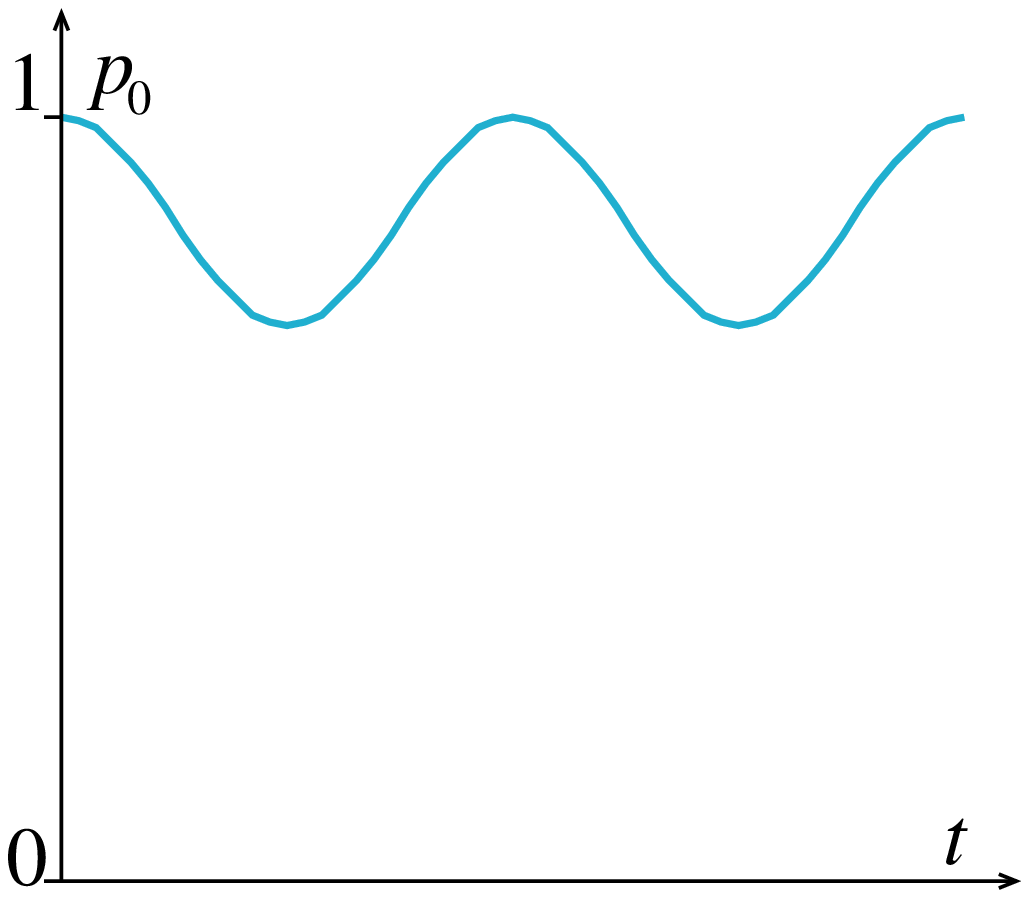}
\end{center}
\caption{a) Evolution for Hamiltonian $h_x\sgm_x+h_z\sgm_z$.
b) Probability $p_0$ for state $\ket{0}$.}\label{Fig:skewrot}
}
\end{figure}

The Hamiltonian proportional to $\sgm_z$ was implemented
by changing depth of one well \Fig{wella} if amplitude of
transition is negligible. On the other hand, the potential should be 
symmetric for application of $\sgm_x$ described above. Undesirable combination
of tunneling and asymmetry is demonstrated on \Fig{skewrot}

The picture \Fig{skewrot} shows, that for system in initial state $\ket{0}$ and
for triple $(h_x,h_y,h_z)$ corresponding to asymmetric potential $h_z \neq 0$,
oscillations do not have complete sweep (unlike $h_z=0$, \Fig{rotx}b) and such 
system never reach state $\ket{1}$.

\begin{figure}[htb]
\figframe{
\begin{center}
a)~\includegraphics[scale=0.5]{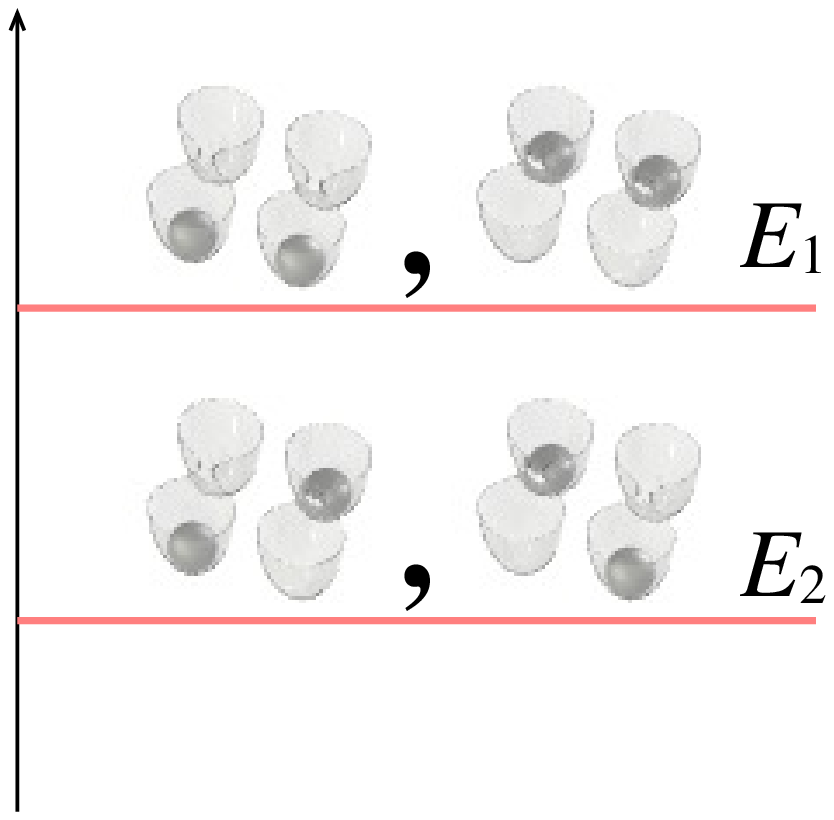}
~~
b)~\includegraphics[scale=0.5]{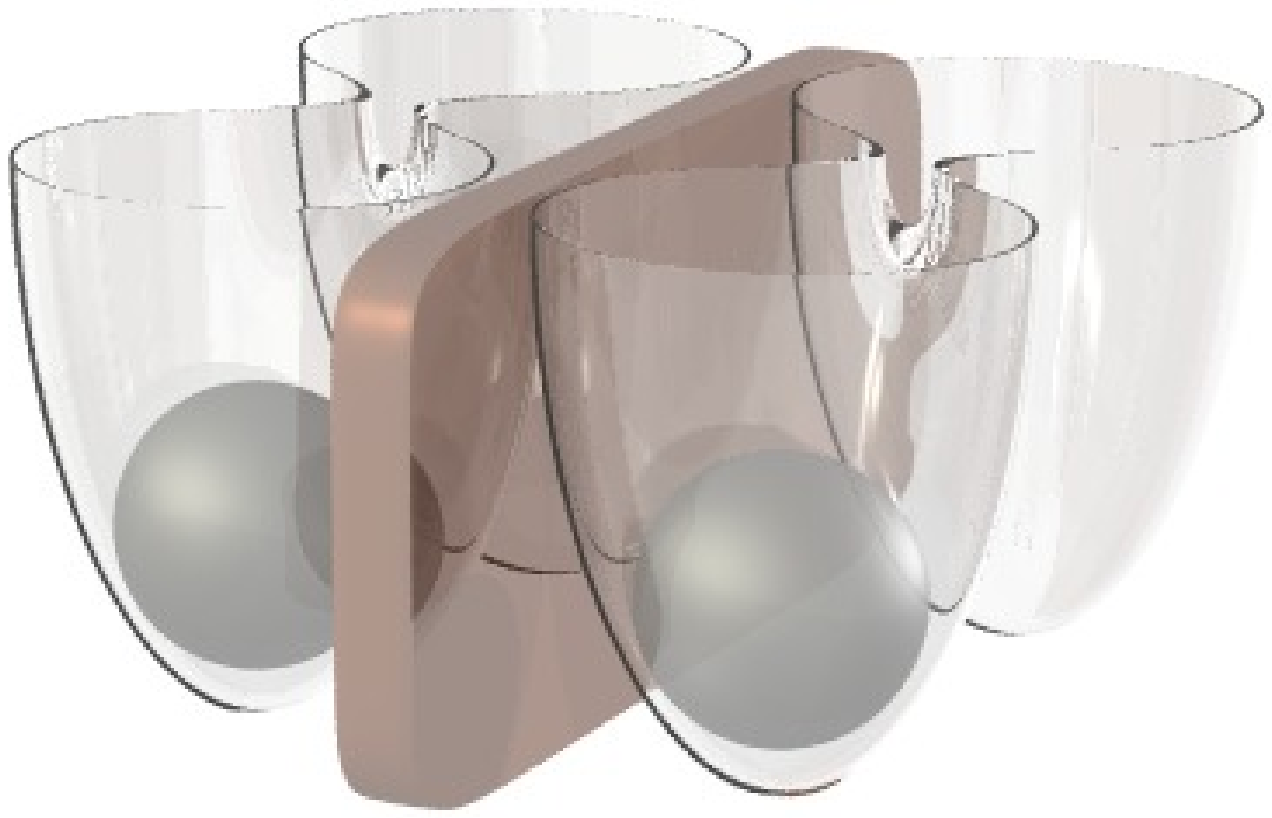}
\end{center}
\caption{a) Energy levels for simple interaction of two double wells: 
$E_1$ for states $\ket{00},\ket{11}$ and $E_2$ for $\ket{01},\ket{10}$. 
b) The quantum wells with screening separator (state $\ket{00}$).}
\label{Fig:well2d}
}
\end{figure}

It was described only one-qubit gates, but for universality it is necessary 
to use two-qubit gates also. In universal set
\Eq{Enew} such gates are generated by Hamiltonian $\sgm_z \otimes \sgm_z$
applied to adjacent qubits. It is convenient, that such Hamiltonian is
diagonal in computation basis. It is enough to consider interaction,
then energy of two systems in same states $\ket{00}$ and $\ket{11}$ 
is not equal to energy in states $\ket{01}$ and $\ket{10}$, \ie\ in
basis $\ket{00}$,$\ket{01}$,$\ket{10}$,$\ket{11}$ the Hamiltonian
of interaction has diagonal form
\begin{equation}
 \op H_{int} = \Mat{llll}{E_{1}&0&0&0\\0&E_{2}&0&0\\
                    0&0&E_{2}&0\\0&0&0&E_{1}},
\label{Hzz}
\end{equation}
Simple example is Coulomb interaction between two particles, where
in \Eq{Hzz} $H_{00}=H_{33}=E_1$, and $H_{11}=H_{22}= E_2$ due to symmetry and
$E_1 \ne E_2$ because distances between particles in such states
are different \Fig{well2d}.
 
More general form of Coulomb interaction 
with symmetrical potentials also could have off-diagonal term 
$c_e(\sgm_x\otimes\sgm_x+\sgm_y\otimes\sgm_y)$ corresponding to 
exchange $\ket{01} \leftrightarrow \ket{10}$, but it is small
for wells with big potential barrier, because in such a case any 
off-diagonal process with change of state is suppressed. 

The exchange would have analogue with F\"oster process in exitonic quantum dots 
\cite{Lovett,qnano}. On the other hand the off-diagonal term often was 
discussed for construction of different entangled states, but considered 
set of quantum gates \Eq{Enew} let us prepare any state using only simpler diagonal
two-qubit gate with Hamiltonian \Eq{Hzz}.

More difficult problem is impossibility to ``turn-off'' the Coulomb
interaction, so on \Fig{qudots} and \Fig{well2d}b the two-qubit gates
are depicted schematically as some kind of variable ``screening'' shields.

There are proposals to use design with all-optical control
of exitonic quantum dots \cite{Lovett,qnano,BIZR}, but it is possible to
discuss simpler scheme, where tuning of interaction between two dots 
performed by change of distance.

\subsection{Polymer model with double chain.}

\begin{figure}[htb]
\figframe{
\begin{center}
\includegraphics{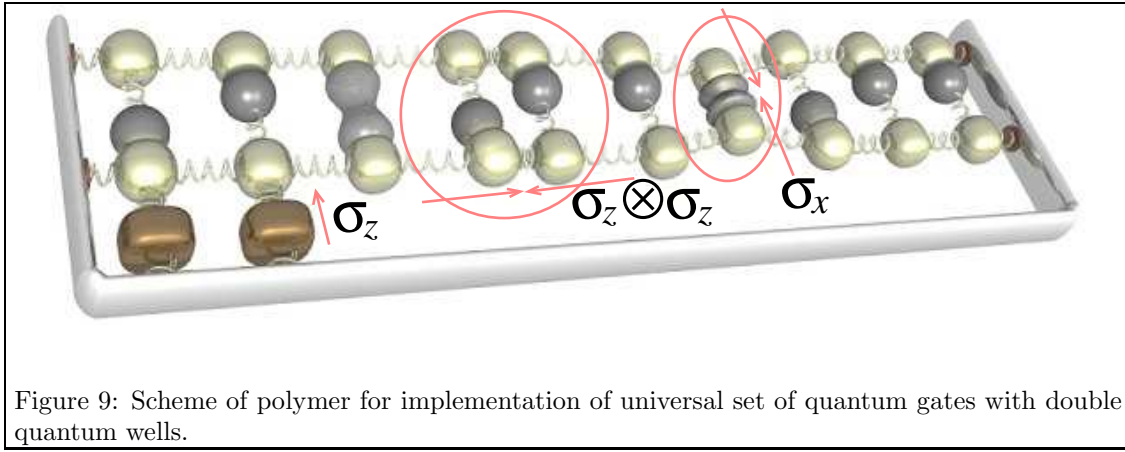} 
\end{center}
\caption{Scheme of polymer for implementation of
universal set of quantum gates with double quantum wells.}\label{Fig:poly}
}
\end{figure}

On \Fig{poly} is depicted scheme of such implementation of quantum
gates with a double polymer chain. As computational basis is
used two states of system in a ``stair,'' it is again a potential well 
with two minima. It is rather abstract scheme may be applied to
different kinds of physical systems, and it should be mentioned,
that different proposals with one-dimensional arrays and polymer chains 
are widely discussed since first papers about practical implementation 
of quantum computer \cite{Lloyd93,Ising94}, till more recent times 
\cite{Benj0,DV0,BLW}.

On \Fig{poly} denoted implementation of all three necessary kinds of gates
from universal set \Eq{Enew}. The gates with Hamiltonian $\sgm^{(k)}_x$ may be
implemented by deformation of polymer double chain with shortening of 
$k$-th ``stair.'' In such a case potential with small tunneling amplitude 
\Fig{welld} is transformed to potential like \Fig{wellb}.

The only necessary two-qubit gates with Hamiltonian 
$\sgm^{(k)}_z \otimes \sgm^{(k+1)}_z$ may be implemented by other kind
of deformation (see \Fig{poly}) with approaching of two ``stairs.'' 
Here is suggested, that tunnelling of particles between two stairs even
in such nearer positions is still impossible for such process and only 
result is change of energy level \Fig{well2d}, because two systems with same state,
\ie\ $\ket{00},\ket{11}$ are closer and interact stronger, than with opposite one 
$\ket{01},\ket{10}$, see \Fig{well2d} and \Fig{poly}.

It was mentioned earlier, that such set with $2n-1$ gates is still not
universal. It is interesting, that for given model the set of gates has
simple meaning as transverse and longitudinal deformations of the double chain. 
The deformations save symmetry of system with respect to reflection, but
two additional gates with Hamiltonian $\sgm_z$ are necessary for universality 
and do not have such property.

In given model such gates are considered as asymmetric deformation of
potential due to additional ``handles'' from one side of ``stair.'' 
The set of quantum gates \Eq{Enew} is convenient, because it is
necessary to apply such operation only to last two systems in array.

\medskip

Maybe such polymer system is not too fine for precise implementation
of quantum gates, but convenient for metrological and complexity 
research. For example due to excitation of transverse and longitudinal
vibrations of polymer may be generated set of gates with only quadratic 
complexity, additional asymmetric interaction with first system still 
does not change the situation and only special actions on {\em two} first 
``stairs'' of chain produce universal set and exponential complexity. 
The example with non-universality shows, that the exponential complexity 
is not an omnipresent elementary property of arbitrary quantum evolution
and may be related with rather fine tuning.

\subsection{Control of chain without access to every qubit}

The same property of considered universal set of quantum gates also provides 
useful possibility to control chain only by action on the ends. Really,
any combination of quantum gates may be decomposed in consequent actions
of pairs: 1) action of gates from subgroup generated by transverse ($\sgm^{(k)}_x$) 
and longitudinal ($\sgm_z^{(j)}\otimes\sgm_z^{(j+1)}$) deformations of 
chain, 2) asymmetric action on two first qubits ($\sgm^{(1)}_z$ and $\sgm^{(2)}_z$).

The access to end of chain is quite simple, but deformations of arbitrary segment
may be difficult operation. To avoid such a problem, let us consider wave of
transverse and longitudinal deformations exited near ends of chain. 
Let us consider \Fig{poly} as demonstration of two waves travelling in opposite
directions. The result of such process depends on site, there such waves met.

Full group generated only by such deformations is Spin$(2n)$, isomorphic
with group SO($2n$) of rotations in 2n-dimensions (up to $\pm 1$ multiplier)
and so it is possible to associate the excitations of chain with rotations. 
The whole set of gates produced by such scenario is universal, if subset of
symmetric wave excitations discussed above is universal in SO$(2n)$ group.

Such analysis depends on concrete form of excitations, but the group SO$(2n)$
is {\em not} exponential like U$(2^n)$, and so it maybe performed in principle.
If we know, how to generate any element of SO$(2n)$ group, then using
combinations of consequent excitations of chain together with asymmetric ($\sgm_z$)
actions on last two segments it is possible to generate any element of U$(2^n)$.

\begin{figure}[htb]
\figframe{
\begin{center}
\includegraphics{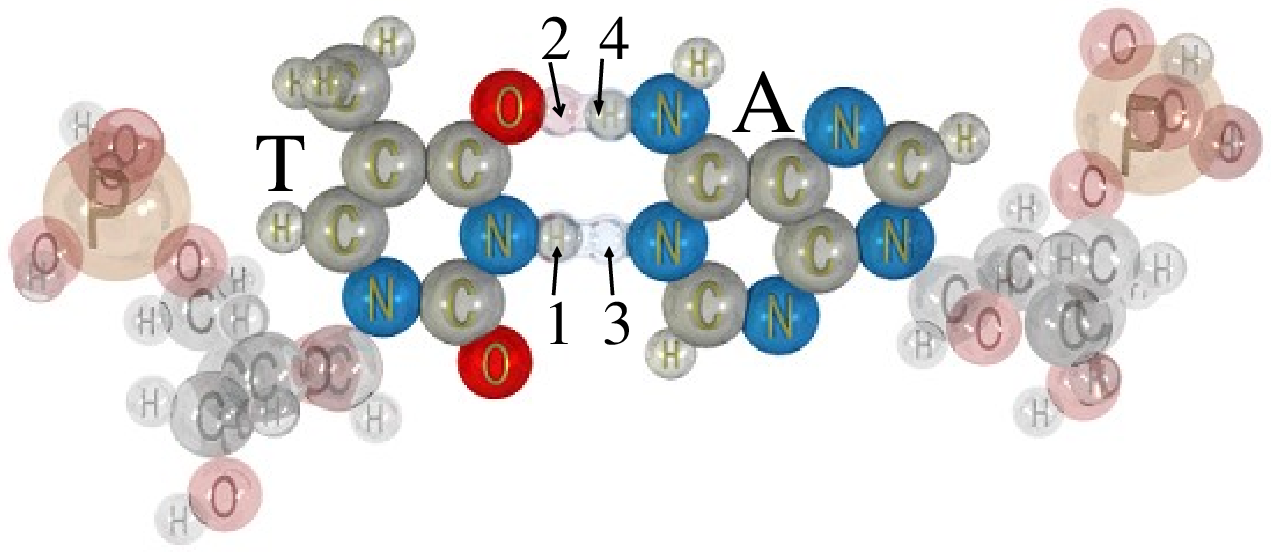} 
\end{center}
\caption{Scheme of tautomeric transitions in DNA base pair. Hydrogen positions 
1 and 4 correspond to usual TA pair. Positions 2 and 3 --- T$_{\text{enol}}$ and 
A$_{\text{imino}}$ forms respectively \cite{Tau,DH}.}\label{Fig:tau}
}
\end{figure}

It should be mentioned also, that more rigor analysis of the model may
demand quantum approach to excitations of chain, \ie\ {\em phonons} and
so becomes closer to some ideas discussed in \cite{Benj0} or \cite{BLW}.
The interaction of vibrational modes with quantum transitions in ``stairs''
(bases) of double helical chain also is important for analysis of mutations in DNA
and a quantum model (with different, bosonic model of transitions) may be
found for example in \cite{Tau}, but structures (see \Fig{tau}) and processes 
are more complex and should be discussed elsewhere.

\subsection{Initialization and measurement}
\subsubsection{Two alternative approaches}

\begin{figure}[htb]
\figframe{
\begin{center}
\includegraphics{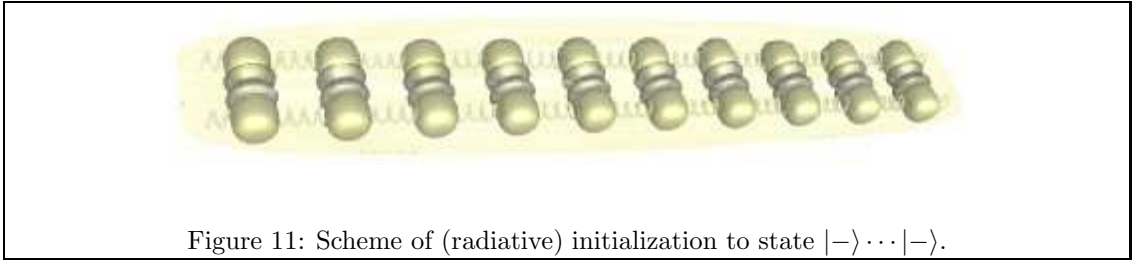} 
\end{center}
\caption{Scheme of (radiative) initialization to state $\ket{-}\cdots\ket{-}$.}
\label{Fig:sqpoly}
}
\end{figure}

Initialization of a chain in some fixed state is necessary for quantum
computations and may be done using different methods. It was discussed above,
that for small value of potential barrier \Fig{wellb} instead of almost
degenerate system there are two states $\ket{-}$ and $\ket{+}$ with 
different energies. So if transition to lower state $\ket{-}$ with emission of
photon is allowed it is enough to set chains in closest position and wait some time%
\footnote{Here formal irreversibility of such process is related with 
consideration of open system, otherwise emitted photons could always be 
absorbed with transition to upper level.}
to initialize system in state $\ket{-}\cdots\ket{-}$ \Fig{sqpoly}.

Schemes of measurement may be similar with used in more traditional double quantum
dots design \cite{GBMes}, but another approach to initialization and measurement
exists also.

\begin{figure}[htb]
\figframe{
\begin{center}
\includegraphics{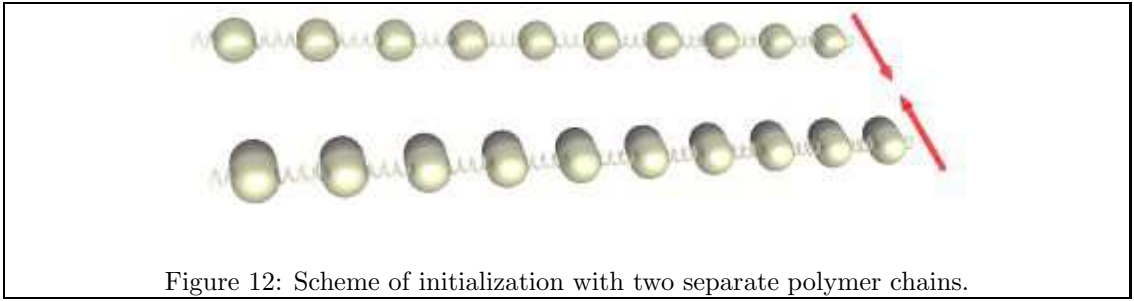} 
\end{center}
\caption{Scheme of initialization with two separate polymer chains.}
\label{Fig:ipoly}
}
\end{figure}

The model uses double polymer chain, so it is possible to prepare
for initialization two separate chains of different kinds and join them \Fig{ipoly}. 
First chain here contains only empty nodes and second one --- only nodes 
with attached system, so double chain after union 
has state $\ket{0}\cdots\ket{0}$ \Fig{ipoly}.

\begin{figure}[htb]
\figframe{
\begin{center}
\includegraphics{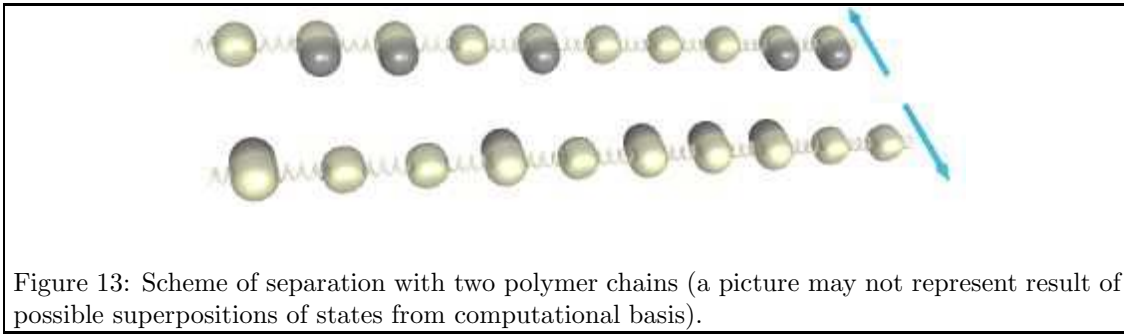} 
\end{center}
\caption{Scheme of separation with two polymer chains (a picture may not 
represent result of possible superpositions of states from computational basis).}
\label{Fig:sepoly}
}
\end{figure}

Similar approach is possible to use for measurement of state.
It is necessary first to disjoin two chains \Fig{sepoly}. Formally
it simply corresponds to infinite big value of separation parameter in
already used model with quantum gates. 

The separation alone is not measurement process, but 
now transitions between chains are not possible, and it is enough to consider 
task of measurement with single polymer chain containing two kinds of nodes.

Such ``destructive'' approach may be too rough for some fine quantum
algorithms, but again is essential for measures of complexity and other
fundamental issues underlying quantum computation and control.
Two chains are joining together (initialization, \Fig{ipoly}), suffering
different kinds of interactions, corresponding to some set of quantum gates 
described above (quantum control, or ``computation'') and finally are separating.
If the system really possesses ``universal access'' to whole exponentially big
Hilbert space of quantum states? If the separated chains display desired kind 
of quantum correlations?

The question about universality was already discussed above. For analysis
of quantum correlations and different measurement problems the considered 
scheme with separated chains \Fig{sepoly} may be also convenient due to
closer resemblance with standard design in most experiments with quantum 
correlations, communications {\em etc.}

\subsubsection{Digression to quantum communications area}

\begin{figure}[htb]
\figframe{
\begin{center}
\includegraphics{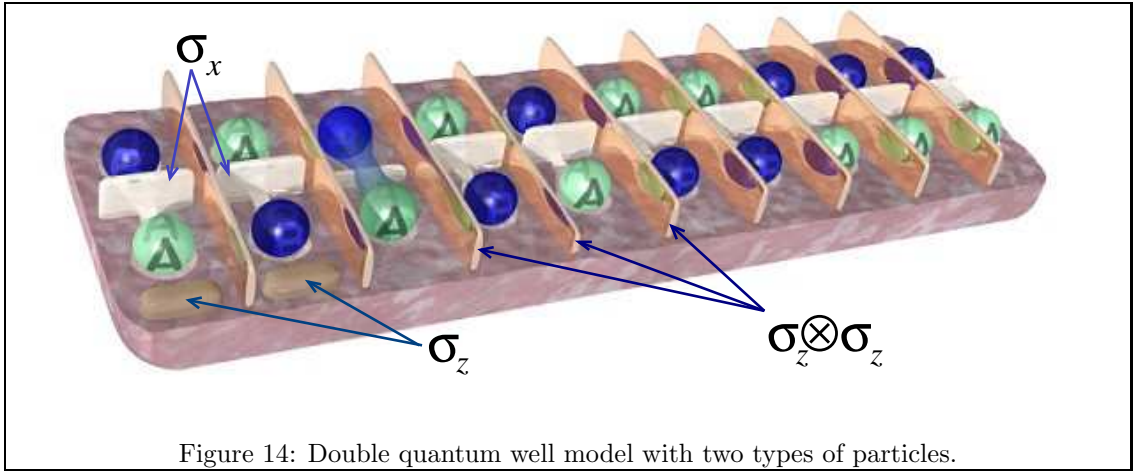} 
\end{center}
\caption{Double quantum well model with two types of particles.}\label{Fig:dqd2}
}
\end{figure}

For more generality it is convenient to consider slighly different
design of double quantum well array with two kinds of systems \Fig{dqd2}.
We have two systems $A$ and $B$ and two states of quantum well
$\ket{0} = \ket{AB}$, $\ket{1} = \ket{BA}$. 

The double chain used above also may be represented in such a way, if
to consider empty node as $A$ and node with attached system as $B$.
For similar situations with $A$ and $B$ are two different systems 
there are two different cases: it is possible to consider quantum superposition
of the systems, or it is prohibited by superselection rule 
(see also \cite{DV0}). And finally $A$ and $B$ may be simply two
different states of the same system, for example two states of photon
with different polarizations.

\begin{figure}[htb]
\figframe{
\begin{center}
\includegraphics{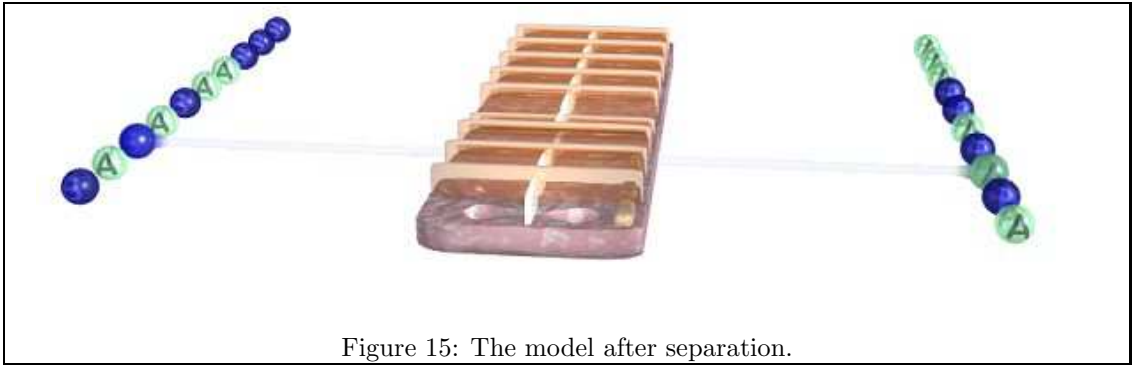} 
\end{center}
\caption{The model after separation.}\label{Fig:dqs2}
}
\end{figure}

Let us consider now, that the systems in array are separated \Fig{dqs2}
and have sent to Alice and to Bob, two usual personages
in the area of quantum communications.

If the systems $A$ and $B$ are two different states of photon, then
it is similar with standard design for testing Bell inequalities, quantum 
communications, cryptography {\em etc.}, \cite{BellUnsp,Asp82,BEZ00}, 
but for more difficult processes with many correlated photons.

If the state of system with $n$ double wells was expressed as sum with $2^n$
terms 
\begin{equation}
\sum_\veK \alpha_\veK \ket{\veK} =
\sum_{k_1,\ldots,k_n}\alpha_{k_1,\ldots,k_n} \ket{k_1}\cdots\ket{k_n}, 
\quad k_j =0,1, \quad j=1,\ldots,n,
\end{equation}
then result of separation may be expressed as
\begin{equation}
 \sum_\veK \alpha_\veK \ket{\veK_A}\ket{\veK_B} =
 \sum_\veK \alpha_\veK \ket{\veK_A}\ket{ \veK_{\neg A}},
\label{alSep}
\end{equation}
where $\neg A$ is bitwise {\sf NOT} and for any $\veK$ state
$\veK_A$ is produced as $\ket{0}\to\ket{A}$, $\ket{1}\to\ket{B}$,
{\em e.g.,} 
$$
\ket{\veK} = \ket{00101}\quad \Longrightarrow\quad
 \ket{\veK_A}=\ket{AABAB},\quad
\ket{\veK_B} = \ket{ \veK_{\neg A}} = \ket{BBABA}.
$$

In a case with $n$ pairs of photons both Alice and Bob would receive $n$ entangled
systems and for large amount of such events they could try to look for quantum 
correlations, like in experiments with Bell's inequalities.
On the other hand, the example with states of same system has a problem 
with undisturbed transmission, due to possible transitions 
between $\ket{A}$ and $\ket{B}$ already after separation and
before measurement.

\medskip

Let us now consider another case, when $A$ and $B$ are two different systems.
From the one hand, such a model may be safer, because only possible error after 
transmission is {\em change of relative phases} for coefficients $\alpha_\veK$ 
in superposition \Eq{alSep}. For any other transformation of state \Eq{alSep} it is
necessary to perform an operation with both parts of separated system, but change 
of phase may be caused by difference of energies for particular states of each chain. 

On the other hand, for such a system {\em there is a problem with distinguishing
between quantum and classical correlations}. It is enough to consider one pair of 
particles after separation
\begin{equation}
 \psi_{AB}= \alpha \ket{\sep{A}{B}} + \beta \ket{\sep{B}{A}}.
\label{MacSup}
\end{equation}
Even if superposition of $A$ and $B$ prohibited by superselection rule,
state \Eq{MacSup} is valid, because it is superposition of two states
of same system $AB$, it is simply ``limiting case'' of double well with
almost zero tunneling amplitude.

Unlike testing Bell inequalities with photons such kinds of superposition 
are still not checked due to enormous difficulty of such experiments. 
It is not clear even, if it is possible in principle to check directly
such superposition --- the problem, that for test of quantum correlations
each party should have possibility not only measure state $\ket{A}$ and
$\ket{B}$, but some superposition of the states $\ket{A}$ and $\ket{B}$
\cite{BellUnsp,Asp82}. 

For example with state of same system it is possible and checked in many
experiments with entangled photons \cite{Asp82}, but for different systems 
superposition of states $\ket{A}$ and $\ket{B}$ may be prohibited by 
superselection rule and usual scheme of testing quantum correlations becomes 
impossible.

Even if superposition is not prohibited by superselection rules,
it may be not allowed by other reasons. Let us consider example,
then both, Alice and Bob are trying to measure systems in
basis $0.5^{0.5}(\ket{A}\pm\ket{B})$, but in such
a case any joint outcome of such measurement may be represented as
superposition 
$
 \frac{1}{2}(\pm\ket{AB}\pm\ket{BA}\pm\ket{AA}\pm\ket{BB}),
$
but terms $\ket{AA}$ and $\ket{BB}$ of such superposition correspond to physically 
impossible states, if initially only one system of each kind presents.

Anyway, existence of some superposition states for joint ``$AB$'' system like
\begin{equation}
 \ket{\opl-_{^{AB}}}=\frac{\sqrt{2}}{2} \bigl(\ket{AB} - \ket{BA}\bigr), \quad
 \ket{\opl+_{^{AB}}}=\frac{\sqrt{2}}{2} \bigl(\ket{AB} + \ket{BA}\bigr),
\label{pmAB}
\end{equation}
is rather usual quantum phenomenon. Often it may be checked measuring 
transitions due to emission or absorption between the energy eigenstates \Eq{pmAB}
like in organic dyes \cite{FLPQ}.
The problem appears during and after separation and may be also explained
using ideas of ``einselection'' theory \cite{ZurEin}.

\begin{figure}[htb]
\figframe{
\begin{center}
\includegraphics{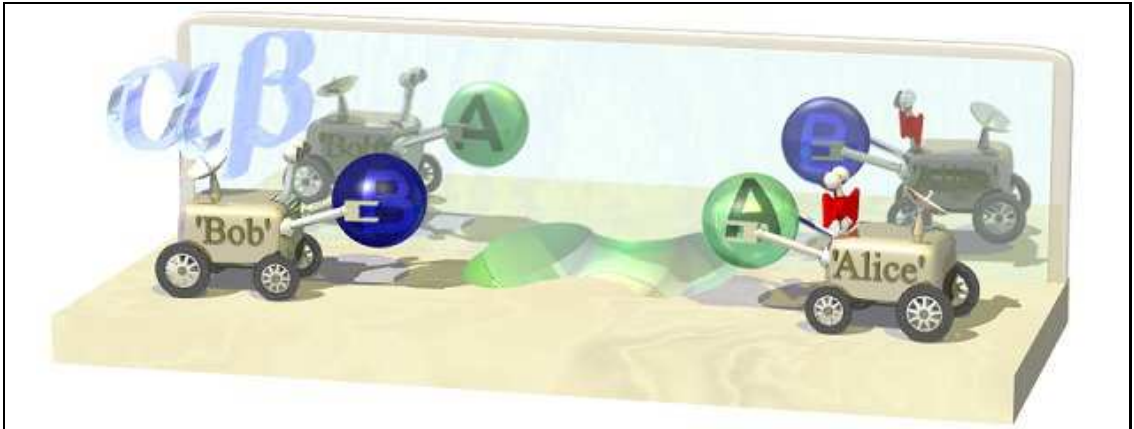}
\end{center}
\caption{State $\alpha\Ket{\mbox{Alice with $A$ and Bob with $B$}}+
    \beta\Ket{\mbox{Alice with $B$ and Bob with $A$}}$. 
 Alice and Bob are depicted  as unspecified ``agents.'' 
 The ``$\beta$-branch'' is shown ``through the looking glass.''} 
\label{Fig:mech}
}
\end{figure}

The only thing we can expect for series of experiments with separation of
states \Eq{pmAB}, that with equal probability will 
be detected two outcomes, \Fig{mech}: 
1) Alice receiving
system A and Bob receiving system B, 2) Alice receiving system B and Bob 
receiving system A. But the same result may be generated by classical
experiment, then system is separated with equal chance either in state 
AB or in state BA. More general case \Eq{MacSup} of superposition with arbitrary 
$\alpha$ and $\beta$ corresponds to probabilities $|\alpha|^2$ and $|\beta|^2$.

In principle, measuring of quantum correlations may be still possible with cooperation
of Alice and Bob. The simplest example is union of spearate systems and measuring
state of $AB$. It is also possible to perform ``conditional exchange'' with
state transfer to auxiliary system:
\begin{equation}
 \bigl(\alpha \ket{\sep{A}{B}} + \beta \ket{\sep{B}{A}}\bigr)\ket{0}
 ~\longrightarrow~
 \ket{\sep{A}{B}}\bigl(\alpha\ket{0}+\beta\ket{1}\bigr).
\label{CondExch}
\end{equation}
The operation \Eq{CondExch} is described by action of {\sf SWAP} gate
\begin{equation}
 \Mat{cccc}{1&0&0&0\\0&0&1&0\\0&1&0&0\\0&0&0&1}
 \quad \mbox{in basis} \quad
 \begin{array}{l}
  \ket{\sep{A}{B}}\ket{0} \\
  \ket{\sep{B}{A}}\ket{0} \\
  \ket{\sep{A}{B}}\ket{1} \\
  \ket{\sep{B}{A}}\ket{1} 
 \end{array}.
\end{equation}
After such operation each participant receive ``proper marble''
($A$ for Alice and $B$ for Bob) and state is stored using auxiliary
local system not suffering from superselection limitations.

Here cooperation between Alice and Bob is essential. {\em E.g.}, in initial
model \Fig{sepoly} $A$ is node of chain and $B$ is node with system
(or otherwise). It is very close to example with $A$ is a system 
and $B$ is ``nothing'', it was simply not convenient in considered model 
to talk about registration of event like ``Bob receives nothing.'' 
On the other hand, it  corresponds to standard interferometer setup 
with two mirrors or two slits experiments. So registration of quantum
correlations is possible, but need for cooperation between parties, 
it corresponds to interference of initially separated beams in the simpler
experiment with one system.

The problem, that in more difficult case with two systems we most likely 
will not be able to measure any correlation. 
It is the usual problem with environmental superselection. If system
had chance to interact with environment, instead of \Eq{CondExch}
it is necessary to write
\begin{equation}
 \bigl(\alpha \ket{e_0}\ket{\sep{A}{B}} + \beta  \ket{e_1}\ket{\sep{B}{A}}\bigr)\ket{0}
 ~\longrightarrow~
 \ket{\sep{A}{B}}\bigl(\alpha\ket{e_0}\ket{0}+\beta\ket{e_1}\ket{1}\bigr),
\label{CondExchEn}
\end{equation}
where $\ket{e_i}$ are states of environment. 
In \Eq{CondExch} auxiliary system had state $\alpha\ket{0}+\beta\ket{1}$ 
and it make possible to measure quantum corellations, but it is not so for 
\Eq{CondExchEn} and $\alpha\ket{e_0}\ket{0}+\beta\ket{e_1}\ket{1}$.

Such environment induced decoherence process often is explained as 
supression of off-diagonal elements of reduced density matrix, but may be 
described using expressions with pure states as well \cite{ZurDar,ZurPen}. 
We still have superposition, considered kind of decoherence is not relevant
with {\em definite outcomes in quantum measurements} \cite{Schl}. On the
other hand, there are different kinds decoherence processes \cite{Lanl7}.

For ``not accurate'' separation it may be produced ``classical kind'' of definite 
outcomes instead of superposition. Such process is possible, because for 
asymmetric potential we have two energy levels (see \Fig{wella} on page 
\pageref{Fig:wella}) and if emission with transition to level with 
lower energy is possible, the system becomes localized in one well. Similar kind
of ``radiative'' decoherence processes was already discussed and used above 
for preparation of initial state (see \Fig{sqpoly} on page \pageref{Fig:sqpoly}).
It should be mentioned also, that after such transition with asymmetric
potential, system may stay localized around one minimum even for further coherent
evolution (see \Fig{skewrot} on page \pageref{Fig:skewrot}).

After separation decoherence processes may not cancel superposition and
only hide and even ``lock'' it, formally it produces some analogue of isolated 
``Everett branches'' \cite{Schl,Ev}. 
The possibility of ``branching'' sometime causes active objections, but except 
of aesthetical reasons there is still not found any strong evidence against it. 
Really idea of quantum computer was considered from very beginning as a possible 
experimental test of Everett's interpretation of quantum mechanics \cite{Deu86}. 
 
\smallskip

Anyway such quantum register is not worse for measurements, than some other 
models, it is rather more convenient for explanation 
of some standard problems. If we may guarantee, that separation is performed
accurate enough, \ie\ does not change coefficients $|\alpha|$ and $|\beta|$
in \Eq{CondExch} and if we may prepare quantum register in same state $\psi$ 
arbitrary number of times, we may use standard methods of quantum tomography
to find the state. It is necessary to have possibility not only measure
the state $\psi$, but also perform some set of known quantum gates 
(say $\sgm_\nu$) before some measurements in the series.
The same is true for arbitrary number of qubits \cite{WootPic}.

\end{document}